\documentclass[twocolumn,prb,showpacs]{revtex4}

\usepackage{graphicx}
\usepackage{dcolumn}
\usepackage{bm}

\begin{document}

\title{Electronic structure properties and BCS
superconductivity in $\beta$-pyrochlore oxides: KOs$_2$O$_6$}

\author{R. Saniz, J. E. Medvedeva, Lin-Hui Ye}
\affiliation{Department of Physics and Astronomy,
Northwestern University, Evanston, Illinois 60208-3112, USA}
\author{T. Shishidou}
\altaffiliation[Now at the ]{Department of Quantum Matter, ADSM,
Hiroshima University, Higashihiroshima 739-8530, Japan.}
\affiliation{Department of Physics and Astronomy,
Northwestern University, Evanston, Illinois 60208-3112, USA}
\author{A. J. Freeman}
\affiliation{Department of Physics and Astronomy,
Northwestern University, Evanston, Illinois 60208-3112, USA}

\date{\today}
                                                                
                                             
\begin{abstract}

We report a first-principles density-functional calculation of the
electronic structure and properties of the recently discovered
superconducting $\beta$-pyrochlore oxide KOs$_2$O$_6$. We find that
the electronic structure near the Fermi energy $E_{\rm F}$ is
dominated by strongly hybridized Os-$5d$ and O-$2p$ states.
A van Hove singularity very close to $E_{\rm F}$ leads to a
relatively large density of states at $E_{\rm F}$, and the Fermi
surface exhibits strong nesting along several directions.
These features could provide the scattering processes
leading to the observed anomalous temperature dependence of the
resistivity and to the rather large specific heat mass enhancement
we obtain from the calculated density of states and the observed
specific heat coefficient. An estimate of $T_c$ within the
framework of the BCS theory of
superconductivity taking into account the possible effects
of spin fluctuations arising from nesting yields the experimental
value.

\end{abstract}

\maketitle

Transition metal (TM) oxides are of intrinsic interest in
physics because of the very rich phenomenology they exhibit due
to electron correlations, ranging from metal-insulator transitions
to colossal magnetoresistance and high critical temperature
superconductivity.  TM oxide compounds with the pyrochlore
structure have long been studied and have found many
applications thanks to their diverse electronic
properties,\cite{subramanian83} but it is not until recently that
superconductivity was found in one such a material, namely
Cd$_2$Re$_2$O$_7$.\cite{hanawa01,sakai01}
Although its superconducting critical temperature
turned out to be low ($T_c\simeq$1 K),
it was an important discovery because
it opened research in this area to a new class of materials.
Very recently, superconductivity was reported
in KOs$_2$O$_6$,\cite{yonezawa04,hiroi04} a so-called
$\beta$-pyrochlore, with a $T_c$ of 9.6 K. More reports
of superconductivity in the same family of compounds have
followed at a rapid pace, with superconductivity being observed
in RbOs$_2$O$_6$ ($T_c=6.3$ K)\cite{yonezawa04b,bruhwiler04} and in
CsOs$_2$O$_6$ ($T_c=3.3$ K),\cite{yonezawa04c} adding to the interest
in these materials.

The discovery of superconductivity in the $\beta$-pyrochlores raises,
of course, the question of the underlying mechanism. While the
mechanism in
Cd$_2$Re$_2$O$_7$, an $\alpha$-pyrochlore,\cite{yonezawa04c}
can be understood within the weak-coupling 
Bardeen, Cooper, and Schrieffer (BCS) theory of
superconductivity,\cite{hiroi02} 
Hiroi and co-workers have suggested\cite{hiroi04}
from the outset that KOs$_2$O$_6$ is an unconventional superconductor,
with the pairing mediated by spin fluctuations.\cite{koda04}
On the other hand, Batlogg and co-workers suggested\cite{bruhwiler04}
that RbOs$_2$O$_6$ could be a conventional BCS-type superconductor,
and recent pressure effects measurements appear to bring further
support to their conclusions.\cite{khasanov04}
Given the close similarity between these two compounds, it seems
unlikely that their
superconductivity has a different origin. Clearly, a careful
study of the electronic structure of these materials may
shed light on the superconductivity mechanism.

In this work, we focus on  KOs$_2$O$_6$ and carry out a
self-consistent
first-principles density-functional calculation of its
electronic structure, using a newly parallelized
implementation\cite{code} of the
full-potential linearized augmented plane wave (FLAPW)
method.\cite{wimmer81} Our calculations are made within the Perdew,
Burke, and Ernzerhof generalized gradient approximation
(GGA)\cite{perdew97} to the exchange-correlation potential
and include the spin-orbit coupling (SOC) term in the Hamiltonian.
Our results show that
the electronic structure near the Fermi energy ($E_{\rm F}$) is
dominated
by strongly hybridized Os-$5d$ and O-$2p$ states. There is a van Hove
singularity (vHS) very close to $E_{\rm F}$, leading to a relatively
large density of states (DOS) at $E_{\rm F}$, and the Fermi surface
shows strong nesting along several directions.
These features could provide the scattering processes
leading to the reported\cite{yonezawa04,hiroi04}
anomalous temperature dependence of the resistivity above $T_c$,
and to the rather large specific heat mass enhancement we obtain
from the calculated DOS and the measured low temperature specific
heat coefficient.\cite{hiroi04} We estimate the $T_c$ within the 
BCS framework and are able to obtain the experimental value if
we take into account the possible effects of spin fluctuations
arising from nesting.

KOs$_2$O$_6$ crystallizes in a cubic structure with
space group $Fd\bar 3m$, and has 18 atoms per unit cell:
two K ($8b$), four Os ($16c$), and twelve O ($48f$).\cite{yonezawa04}
The structure has an internal
parameter $x$, that fixes the position of the oxygen atoms.
The experimental lattice constant recently given by Yonezawa
and co-workers is $a=10.101$ {\AA},\cite{hiroi04}
but no value for the $x$ parameter has yet been reported.
We have optimized variationally both the lattice constant and the
internal parameter,
obtaining $a=10.298$ {\AA}, which differs by 1.95\% from the 
experimental value, and $x=0.316$.
The latter can be compared with the value of $x=0.315$ 
recently reported for the related superconducting pyrochlore
RbOs$_2$O$_6$ by Br\"uhwiler and collaborators.\cite{bruhwiler04}
Each Os atom is octahedrally coordinated by six O atoms, with
O-Os-O angles of 88.66$^\circ$ and 91.34$^\circ$ (compare with
88.85$^\circ$ and 91.15$^\circ$, respectively, for RbOs$_2$O$_6$,
reported in Ref.~\onlinecite{bruhwiler04}).
Unlike the perovskite superconductors, in the pyrochlore oxides the
transition metal-oxygen octahedra are not distorted.
The Os-O distance we find is 1.94 {\AA} (compare with,
for example, the calculated\cite{oguchi95,singh95}
Ru-O distances of 1.93 {\AA} and 2.061 {\AA} in Sr$_2$RuO$_4$).
In KOs$_2$O$_6$, the
pyrochlore lattice is formed by highly interconnected Os-O
staggered chains,
resulting in corner sharing tetrahedra, with the Os
ions occupying the vertices.
In Ref.~\onlinecite{bruhwiler04}, it is suggested that
in the $\beta$-pyrochlore oxides, the
transition TM-O-TM
angle plays a role in defining the
superconducting properties,\cite{bruhwiler04} with smaller angles
favoring higher critical temperatures. For KOs$_2$O$_6$, we
find an Os-O-Os angle of 139.16$^\circ$, which is indeed 
smaller than the reported angle of
139.4$^\circ$ for RbOs$_2$O$_6$.\cite{bruhwiler04}
There is currently not enough experimental information regarding
CsOs$_2$O$_6$ for comparison with this material.

\begin{figure}
\includegraphics[width=\hsize]{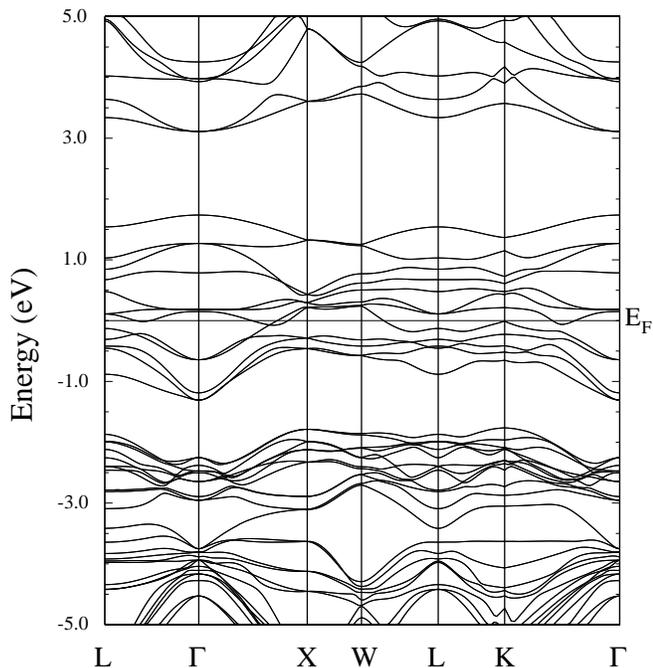}
\caption{\label{fig1} Band structure of KOs$_2$O$_6$, calculated
within the GGA for exchange and correlation and taking into
account spin-orbit coupling self-consistently. 
The twelve bands around $E_{\rm F}$ arise from Os $5d$
and O $2p$ states. A saddle point near the
center of the $\Gamma L$ line causes a vHS
very close to $E_{\rm F}$.}
\end{figure}

\begin{figure}
\includegraphics[width=0.95\hsize]{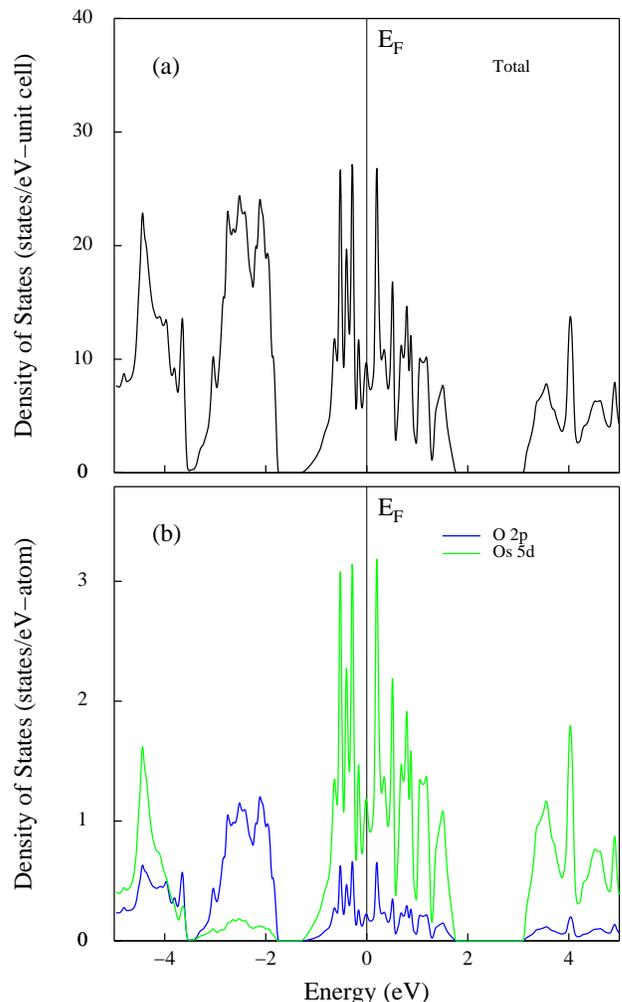}
\caption{\label{fig2} Density of states within the GGA and taking
into account spin-orbit coupling self-consistently. (a) Total DOS,
showing a sharp peak very close to $E_{\rm F}$.
(b) Projected DOS of the O $2p$ and Os $5d$ states, on a per atom
basis.}
\end{figure}

The calculated energy bands along high symmetry directions in
{\bf k}-space, within 5 eV from $E_{\rm F}$, are shown
in Fig.~\ref{fig1}.
There is a manifold of twelve states in the vicinity of
$E_{\rm F}$, with a bandwidth of 3 eV,
separated by relatively large energy gaps above and below.
Two bands cross the Fermi energy:
the lower band cuts the $\Gamma X$
and $WL$ lines and gives rise to a hole-like Fermi surface sheet
in the form of a tubular network; the upper band
crosses $E_{\rm F}$ twice within the first Brillouin
zone (cf. the $\Gamma L$, $\Gamma X$, and $\Gamma K$ lines),
giving rise to two Fermi surface sheets.
As will be illustrated below, this results in an electron-like
closed shell centered at the $\Gamma$ point.
Of importance is the existence of a vHS very close
to $E_{\rm F}$, caused by a saddle point between the two sheets
at $\sim 0.015$ eV below the Fermi level,
near the middle of the $\Gamma L$ line.

\begin{figure}
\includegraphics[width=\hsize]{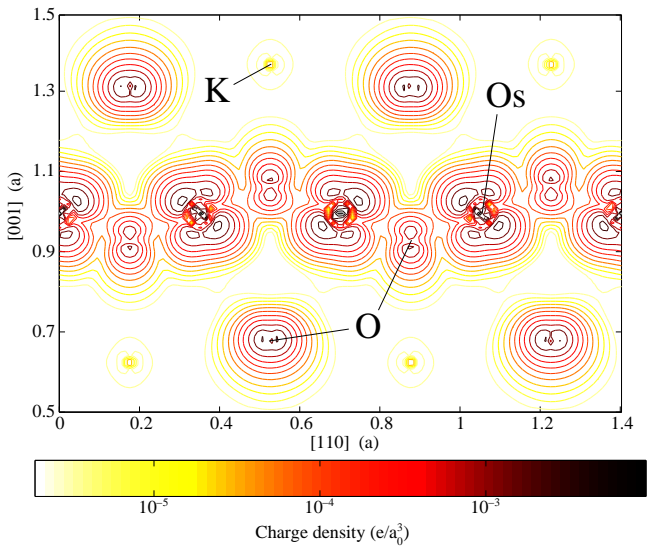}
\caption{\label{fig3} Contour plot of the charge density in the
$(1\bar 10)$ plane for states
lying within 0.027 eV below $E_{\rm F}$. A staggered Os-O chain is
clearly seen. The other oxygen ions in the plot belong to 
chains traversing the $(1\bar 10)$ plane.
As expected, the charge around the potassium ions is extremely
weak.}
\end{figure}

\begin{figure}
\includegraphics[width=\hsize]{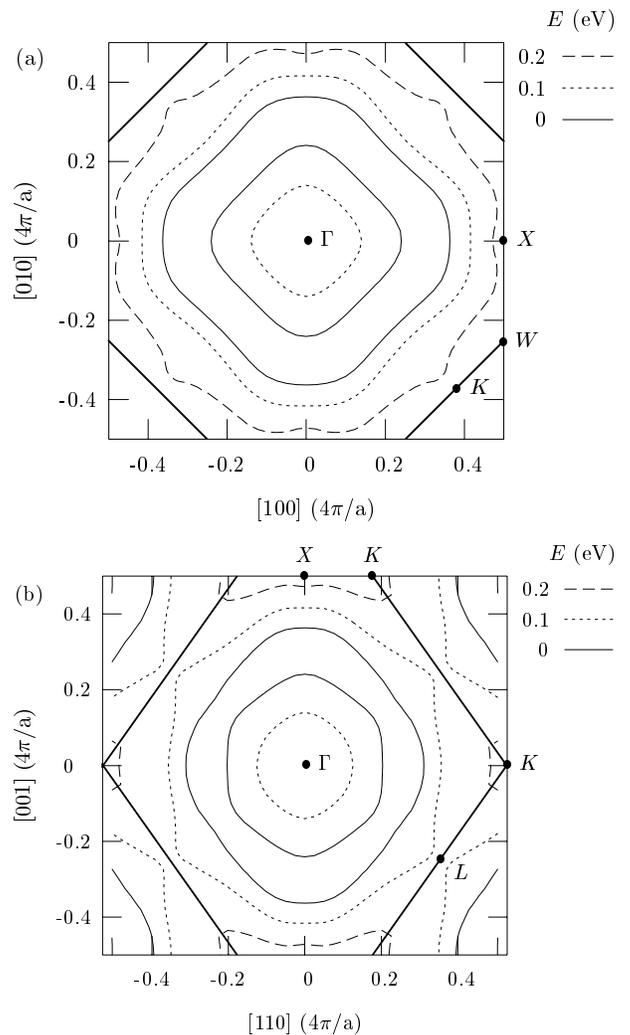}
\caption{\label{fig4} Contour plots of the upper band crossing
$E_{\rm F}$ (cf. Fig.~\ref{fig1}) along two high symmetry planes.
$E=0$ corresponds to $E_{\rm F}$. The thick lines indicate the
first Brillouin zone boundaries. (a) Plot for 
{\bf k} in the $X\Gamma X$ plane. Extensive nesting occurs
for {\bf k} along $\Gamma K$, with $k\simeq 0.4$ and 0.62 $4\pi/a$.
(b) Plot for {\bf k} in the $\Gamma X K$
plane. Strong nesting occurs, e.g., for {\bf k} along $\Gamma L$,
with $k\simeq 0.64\;4\pi/a$.}
\end{figure}

The character of the bands near $E_{\rm F}$ is further analyzed by
examining the density of states (DOS). The total DOS, shown in
Fig.~\ref{fig2}(a),
exhibits a peak very close to $E_{\rm F}$, due to the
vHS mentioned above.
We point out that a precursor of the peak is already present,
at a slightly lower energy,
in a calculation without including spin-orbit coupling;
the inclusion of the latter splits this peak, causes the saddle
point, and pushes the higher peak closer to $E_{\rm F}$.
In Fig.~\ref{fig2}(b) we show
the partial DOS of the O $2p$ and Os $5d$ states, which are by far
the dominant states in the energy range shown.
As in previously studied oxide
superconductors,\cite{oguchi95,singh95,singh02} the TM and O
states are strongly hybridized near $E_{\rm F}$. 
The orbital character of the states near $E_{\rm F}$ is
Os $d\epsilon(xy,yz,zx)$
and O $p\pi$. This is clearly shown in Fig.~\ref{fig3}, which
presents a contour plot of the charge density in the $(1\bar 10)$
plane for states lying within 0.027 eV below
$E_{\rm F}$.\cite{comment1}

An important finding is that the Fermi surface shows strong
nesting, in particular the shell-like sheets.
Contour plots of the eigenenergies
for the corresponding band along two different planes
centered at the $\Gamma$ point
are given in Fig.~\ref{fig4}. Figure \ref{fig4}(a) shows
the contour plot for {\bf k} in the $X\Gamma X$
plane, with strong nesting occurring for
$k\simeq 0.4\; 4\pi/a$ and $k\simeq 0.62\; 4\pi/a$
along the $\Gamma K$ directions. Similarly, Fig.~\ref{fig4}(b) shows
a contour plot for {\bf k} in the $X\Gamma K$ plane,
also showing strong nesting, in particular
for $k\simeq 0.64\; 4\pi/a$ along $\Gamma L$
directions.

Consider further some of the properties deduced from the
electronic structure in relation to experiment. Firstly,
the DOS at the Fermi level is relatively high, $N(E_{\rm F})=9.8$
states/eV-unit cell, yielding a Sommerfeld electronic specific heat
coefficient $\gamma=5.78$ mJ/K$^2$ mol-Os. This
is much lower than the experimental
value of 19 mJ/K$^2$ mol-Os found by Hiroi and
collaborators,\cite{hiroi04} based on their estimation of the
specific heat jump at $T_c$ and the BCS weak-coupling
relation $\Delta C/\gamma T_c=1.43$. Thus, we find a very large
specific heat mass enhancement
$\gamma_{\rm exp}/\gamma_{\rm band}\simeq 3.3$,
though it is still smaller than that of, e.g.,
Sr$_2$RuO$_4$, found to be 3.8 or more.\cite{oguchi95,singh95}
The above mass enhancement represents
a rather high coupling constant, $\lambda=2.3.$
Given the relatively low $T_c$ of KOs$_2$O$_6$,
it appears that, besides electron-phonon, there is a very
important electronic many-body contribution
to the specific heat mass enhancement.
Interestingly, Hiroi and co-workers noted in particular
the unusual behavior of the measured
resistivity as a function of temperature.\cite{yonezawa04,hiroi04}
Data corresponding to single crystals are greatly desirable to
advance further in this direction. Let us just point out here
that, under appropriate
conditions, deviations from conventional Fermi liquid behavior
in two-dimensional systems can be caused by both vHS's
near the Fermi energy\cite{newns94} and by Fermi
surface nesting,\cite{virosztek90} due to the increased phase
space available for electron-electron scattering.\cite{comment2}

From the measured $T_c=9.6$ K, and using $\Delta=1.76 k_{\rm B}T_c$,
the superconducting gap is calculated to be  $\Delta=1.456$ meV.
A calculation of the average Fermi velocity then allows us to
readily estimate the BCS-Pippard coherence length
$\xi_0=\hbar\langle v_{\rm F}\rangle/\pi\Delta$.
We find $\langle v_{\rm F}\rangle=1.47 \times 10^7$ cm/s, 
which yields $\xi_0=212$ {\AA}. This is
an order of magnitude larger than the reported
Ginzburg-Landau coherence length $\xi=30$ {\AA},
obtained from the estimated 
upper critical field $H_{c2}$.\cite{hiroi04,comment3}
Thus, the so-called dirty limit would apply.
Indeed, writing $\xi^{-1}=
\xi_0^{-1}+l^{-1}$, with $l$ the mean free path, yields an estimated
value of $l=35$ \AA. This relatively short value suggests again that
strong scattering processes play an important role in the
electronic properties of this material.

We estimate
the McMillan-Hopfield\cite{mcmillan68,hopfield69}
electron-phonon coupling constant $\lambda_{\rm ep}$, for which the
constituent-weighted average can be written
$\lambda_{\rm ep}=
\sum_i w_i\eta_i/M_i\langle\omega^2\rangle$.\cite{salunke97}
Here, the spherically averaged
Hopfield parameters, $\eta$, were calculated
in the crude rigid-muffin-tin approximation, with\cite{skriver85}
\begin{equation}
\eta=2N(E_{\rm F})\sum_l(l+1)M^2_{l,l+1}
{{f_lf_{l+1}}\over{(2l+1)(2l+3)}},
\end{equation}
where $f_l=N_l(E_{\rm F})/N(E_{\rm F})$ is a relative partial DOS and
$M_{l,l+1}=-\phi_l\phi_{l+1}[(D_l-l)(D_{l+1}+l+2)+
(E_{\rm F}-V)R^2]$ is an electron-phonon matrix element. The
quantities entering the latter are the logarithmic derivatives
($D_l$) and the partial wave amplitudes ($\phi_l$), both evaluated at
$E_{\rm F}$ and at the muffin-tin
radius ($R$); $V$ is the one-electron potential at $R$.
The average phonon frequency can be estimated by
$\langle\omega^2\rangle^{1/2}=0.69\,\Theta_{\rm D}$. There is currently
no reported $\Theta_{\rm D}$ for KOs$_2$O$_6$, but
Br\"uhwiler and co-workers did such measurements for
RbOs$_2$O$_6$.\cite{bruhwiler04}
Under the assumption that the values of $\Theta_{\rm D}$ fall in the
same range for both materials,
we take $\Theta_{\rm D}=240$ K, and obtain $\lambda_{\rm ep}=1.4$.
As expected, this value is well below the total $\lambda$ obtained
from the specific heat mass enhancement discussed above.\cite{comment4}

To estimate $T_c$,\cite{comment5} we use
the Allen and Dynes modification of the
McMillan equation,\cite{mcmillan68,allen75}
including an effective electron-spin interaction coupling
constant $\mu_{\rm sp}$,\cite{bennemann72} i.e.,
\begin{equation}
T_c={{\langle\omega^2\rangle^{1/2}}\over{1.2}}
\exp\left[-{{1.04(1+\lambda_{\rm ep}+\mu_{\rm sp})}\over
{\lambda_{\rm ep}-(\mu^*+\mu_{\rm sp})(1+
0.62\lambda_{\rm ep})}}\right].
\end{equation}
The so-called Coulomb pseudopotential $\mu^*$
can be estimated from the DOS at
$E_{\rm F}$,\cite{bennemann72} and in our case is $\mu^*=0.09$.
On the other hand, there is no simple expression for $\mu_{\rm sp}$.
If $\mu_{\rm sp}=0$ we find $T_c=18$ K.
Strong spin fluctuations,\cite{comment6}
with a coupling constant $\mu_{\rm sp}=0.13$
would yield the experimental value $T_c=9.6$ K. As an estimate,
from our DOS we have calculated the Pauli paramagnetic susceptibility,
finding $\chi_{\rm band}=1.58\times 10^{-4}$ cm$^3$/mol.
From Ref.~\onlinecite{hiroi04}, we estimate the experimental value
at $T=0$ to be
$\chi_{\rm exp}\simeq 4\times 10^{-4}$ cm$^3$/mol,
which yields an enhancement ratio
$\chi_{\rm exp}/\chi_{\rm band}=2.53$. 
Thus, strong spin
fluctuations may indeed play an important role due to the significant
nesting of the Fermi surface.

We are grateful to O. Kontsevoi, M. Weinert, T. Jarlborg, and
J. B. Ketterson
for helpful discussions. This work was supported by the Department of
Energy (under grant No. DE-FG02-88ER 45372/A021
and a computer time grant at the
National Energy Research Scientific Computing Center).


\begin{thebibliography}{99}

\bibitem{subramanian83} M. A. Subramanian, G. Aravamudan,
and G. V. S. Rao, Prog. Solid St. Chem. {\bf 15},
55 (1983).

\bibitem{hanawa01} M. Hanawa, Y. Muraoka, T. Tayama, T. Sakakibara,
J. Yamaura, and Z. Hiroi, Phys. Rev. Lett. {\bf 87} 187001 (2001).

\bibitem{sakai01} H. Sakai, K. Yoshimura, H. Ohno, H. Kato,
S. Kambe, R. E. Walsted, T. D. Matsuda, Y. Haga, and Y. \=Onuki,
J. Phys.: Condens. Matter {\bf 13}, L785 (2001).

\bibitem{yonezawa04} S. Yonezawa, Y. Muraoka, Y. Matsushita,
and Z. Hiroi, J. Phys.: Condens. Matter {\bf 16}, L9 (2004).

\bibitem{hiroi04} Z. Hiroi, S. Yonezawa, and Y. Muraoka,
cond-mat/0402006v1.

\bibitem{yonezawa04b} S. Yonezawa, Y. Muraoka, Y. Matsushita,
and Z. Hiroi, J. Phys. Soc. Jpn. {\bf 73}, 819 (2004).

\bibitem{koda04} A. Koda, W. Higemoto, K. Ohishi, S. R. Saha,
R. Kadono, S. Yonezawa, Y. Muraoka, and Z. Hiroi, cond-mat/0402400.

\bibitem{bruhwiler04} M. Br\"uhwiler, S. M. Kazakov, N. D. Zhigadlo,
J. Karpinksi,
and B. Batlogg, cond-mat/0403526; S. M.Kazakov, N. D. Zhigadlo,
M. Br\"uhwiler,
B. Batlogg, and J. Karpinksi, cond-mat/0403588.

\bibitem{khasanov04} R. Khasanov, D. G. Eshchenko, J. Karpinski,
S. M. Kazakov, N. D. Zhigadlo, R. Br\"utsch, D. Gavillet, and
H. Keller, cond-mat 0404542.

\bibitem{yonezawa04c} S.Yonezawa, Y.Muraoka, Z. Hiroi,
cond-mat/0404220.

\bibitem{hiroi02} Z. Hiroi and M. Hanawa, J. Phys. Chem. Solids
{\bf 63}, 1021 (2002).

\bibitem{code} T. Shishidou, L.-H. Ye, M. Weinert, and A. J. Freeman
(unpublished).

\bibitem{wimmer81} E. Wimmer, H. Krakauer, M. Weinert, and
A. J. Freeman, Phys. Rev. B {\bf 24}, 864 (1981); M. Weinert,
E. Wimmer, and A. J. Freeman, {\it ibid.} {\bf 26}, 4571 (1982).

\bibitem{perdew97} J. P. Perdew, K. Burke, and M. Ernzerhof,
Phys. Rev. Lett. {\bf 77}, 3865 (1996).

\bibitem{oguchi95} T. Oguchi, Phys. Rev. B {\bf 51}, 1385 (1995).

\bibitem{singh95} D. J. Singh, Phys. Rev. B {\bf 52}, 1358 (1995).

\bibitem{singh02} D. J. Singh, P. Blaha, K. Schwarz, and J. O. Sofo,
Phys. Rev. B {\bf 65} 155109 (2002).

\bibitem{comment1} This energy corresponds to 313 K, which is of
the order of a typical $\Theta_{\rm D}$ in metallic pyrochlores;
see Ref.~\onlinecite{bruhwiler04}.

\bibitem{newns94} D. M. Newns, C. C. Tsuei, R. P. Huebener,
P. J. M. van Bentum, P. C. Pattnaik, and C. C. Chi, 
Phys. Rev. Lett. {\bf 73}, 1695 (1994).

\bibitem{virosztek90} A. Virosztek and J. Ruvalds, Phys. Rev. B
{\bf 42}, 4064 (1990).

\bibitem{comment2} Moreover, disorder and/or impurities can play
a critical role in this respect, as shown by
A. Rosch, Phys. Rev. Lett. {\bf 82}, 4280 (1999).

\bibitem{comment3} We point out that although the estimated
$H_{c2}$ for KOs$_2$O$_6$ is high (38.3 T) and more than
twice its Chandrasekhar-Clogston limit $H_p$, the values found
are not without precedent among BCS-superconductors, being
close to those of several ternary molybdenum chalcogenides; see
M. Decroux and \O. Fischer, in
{\it Superconductivity in Ternary Compounds}, edited by M. B. Maple
and \O. Fischer (Springer-Verlag, Berlin, 1982). On the other
hand, our results indicate that KOs$_2$O$_6$ may be a
strong-coupling superconductor, so that some of the values reported
in Ref.~\onlinecite{hiroi04} may require revision; see
D. Rainer and G. Bergmann, J. Low Temp. Phys. {\bf 14}, 501 (1974).

\bibitem{mcmillan68} W. L. McMillan, Phys. Rev. {\bf 167}, 331 (1968).

\bibitem{hopfield69} J. J. Hopfield, Phys. Rev. {\bf 186}, 443 (1969).

\bibitem{salunke97} H. G. Salunke, R. Mittal, G. P. Das, and
S. L. Chaplot, J. Phys.: Condens. Matter {\bf 9}, 10137 (1997).

\bibitem{skriver85} H. L. Skriver and I. Mertig,
Phys. Rev. B {\bf 32}, 4431 (1985).

\bibitem{comment4} Since $\eta$
depends directly on the
$l$-projected DOS at $E_{\rm F}$ for the different atoms, the role of
potassium in this regard is only indirect,
mainly through its participation
in the determination of the structural parameters of the compound.

\bibitem{comment5} A more rigorous estimation of $\lambda$
and of $T_c$
may require a more elaborate approach than the one adopted
here, due to the peak in the DOS
very close to $E_{\rm F}$. See, e.g., E. Cappelluti and L. Pietronero,
Phys. Rev. B {\bf 53}, 932 (1996).

\bibitem{allen75} P. B. Allen and R. C. Dynes, Phys. Rev. B {\bf 12},
905 (1975).

\bibitem{bennemann72} K. Bennemann and J. Garland, AIP Conf.
Proc. {\bf 4}, 103 (1972).

\bibitem{comment6} In transition metals $\mu_{\rm sp}$ is generally
around $\sim 0.05$.
However, for Sc one has $\mu_{\rm sp}=0.15$; see
Ref.~\onlinecite{bennemann72}.

\end{thebibliography}
\end{document}